\documentstyle[12pt]{article}

\newcommand{\be}{\begin{equation}}
\newcommand{\ee}{\end{equation}}
\newcommand{\ba}{\begin{eqnarray}}
\newcommand{\ea}{\end{eqnarray}}
\newcommand{\no}{\nonumber\\}

\newcommand{\two}{\mbox{\boldmath $2$}}
\newcommand{\one}{\mbox{\boldmath $1$}}

\title{A new type of spontaneous CP breaking}

\author{L.\ Lavoura \\
\small Universidade T\'ecnica de Lisboa \\
\small CFIF, Edif\'\i cio Ci\^encia, Instituto Superior T\'ecnico,
\small P-1096 Lisboa Codex, Portugal}

\begin{document}

\maketitle

\begin{abstract}
I present a new form of spontaneous CP violation in which,
in analogy with the left-right-symmetric model,
CP breaking results from the inequality
of two real vacuum expectation values.
In my model there is no scalar-pseudoscalar mixing,
and the smallness of strong CP violation finds a natural explanation.
\end{abstract}

There exist in the literature models of spontaneous P breaking
and models of spontaneous CP breaking.
Amazingly,
the models of spontaneous P breaking
are very different from the models of spontaneous CP breaking.
The usual model of spontaneous P breaking
is the left-right-symmetric model \cite{LRS}.
In it,
the standard-model gauge group SU(2)$_L$ is duplicated
with a new gauge group SU(2)$_R$,
which acts on the right-handed fermions.
The idea is that there are right-handed charged weak interactions,
but they are mediated by gauge bosons $W_R^\pm$
which are much heavier than the known $W^\pm$.
The basic models of spontaneous CP breaking \cite{lee,weinberg}
are utterly different.
In them,
no extra gauge group is introduced,
and there are no extra weak interactions,
CP-symmetric of the usual ones,
mediated by heavy gauge bosons.
Rather,
spontaneous CP breaking hinges on  a clash
between different terms in the Higgs potential
which are sensitive to the same vacuum phase.
As a consequence,
that vacuum phase does not vanish,
generating CP violation.

A typical feature of models of spontaneous CP violation
is scalar-pseudoscalar mixing \cite{ma}.
This consists in the existence of neutral spin-0 particles
which simultaneously have scalar Yukawa couplings
(with unit matrix in Dirac space)
and pseudoscalar Yukawa couplings
(with $i \gamma_5$ matrix in Dirac space)
with the fermions.
(Scalar-pseudoscalar mixing also has signatures
in the gauge interactions and self-interactions
of the neutral spin-0 particles \cite{silva}.)
This has a problematic consequence:
the one-loop-level generation of electric dipole moments
for the fermions
\cite{barger},
which must be carefully controlled or tuned down.

In this paper I suggest a different kind of model
of spontaneous CP violation.
In it,
there are two different charged weak interactions which,
in the limit of unbroken CP symmetry,
have the same strength but are CP-inverted relative to each other.
My model avoids scalar-pseudoscalar mixing
and gives a rationale for the smallness of strong CP violation.
A feature of the model,
which however is avoidable by the introduction
of an extra discrete symmetry,
is the mixing of the usual left-handed fermions
with new ones with different quantum numbers.
From that mixing flavour-changing neutral currents arise.

My model is inspired on a specific left-right-symmetric model
\cite{BM}
in which the usual scalar bi-doublet
was eliminated in favour of singlet fermions.
That model also provided a solution for the strong-CP problem,
and was later generalized \cite{prl}
to a family of models with similar features.

The gauge group is SU(2)$_A \otimes$SU(2)$_B \otimes$U(1).
There is CP symmetry at the Lagrangian level
except for a soft CP-breaking term.
CP symmetry eliminates the CP-violating term
$\theta / (32 \pi^2) \tilde{G} G$
from the strong-interaction Lagrangian.
CP symmetry interchanges SU(2)$_A$ with SU(2)$_B$
and inverts the sign of the hypercharge.
The gauge coupling constants of SU(2)$_A$ and of SU(2)$_B$
are equal as a consequence of CP;
they are named $g$.
The gauge coupling constant of U(1) is $g^\prime$.
There are two scalar doublets,
\be
\phi_A: (\two, \one)_{1/2} \ \mbox{\rm and} \
\phi_B: (\one, \two)_{1/2}.
\ee
Their lower components have vacuum expectacion values $v_A$ and $v_B$,
respectively.
$v_A$ and $v_B$ are real without loss of generality.
The electric-charge generator is $Q = T_{3A} + T_{3B} + Y$.
I define as usual the conjugate doublets
\be
\tilde{\phi}_A: (\two, \one)_{-1/2} \ \mbox{\rm and} \
\tilde{\phi}_B: (\one, \two)_{-1/2}.
\ee
These doublets are written
\[
\phi_A = \left( \begin{array}{c}
\varphi_A^+ \\ v_A + (H_A + i \chi_A) / \sqrt{2}
\end{array} \right), \
\phi_B = \left( \begin{array}{c}
\varphi_B^+ \\ v_B + (H_B + i \chi_B) / \sqrt{2}
\end{array} \right),
\nonumber
\]
\be
\tilde{\phi}_A = \left( \begin{array}{c}
v_A + (H_A - i \chi_A) / \sqrt{2} \\ - \varphi_A^-
\end{array} \right), \
\tilde{\phi}_B = \left( \begin{array}{c}
v_B + (H_B - i \chi_B) / \sqrt{2} \\ - \varphi_B^-
\end{array} \right).
\ee
The fields $\varphi_A^\pm$ and $\varphi_B^\pm$ are Goldstone bosons,
to be absorbed in the longitudinal components
of $W_A^\pm$ and $W_B^\pm$,
respectively.
$\chi_A$ and $\chi_B$ are neutral Goldstone bosons.
There are two physical neutral scalars,
which are orthogonal combinations of $H_A$ and $H_B$.

Under CP
\be
\phi_A \rightarrow \phi_B^\ast \ \mbox{\rm and} \
\phi_B \rightarrow \phi_A^\ast.
\ee
The scalar potential is
\ba
V &=& \mu_A \phi_A^\dagger \phi_A + \mu_B \phi_B^\dagger \phi_B
\no
& & + \lambda_1
\left[
\left( \phi_A^\dagger \phi_A \right)^2
+ \left( \phi_B^\dagger \phi_B \right)^2
\right]
\no
& & + \lambda_2
\left(
\phi_A^\dagger \phi_A \right) \left( \phi_B^\dagger \phi_B
\right).
\ea
It is CP-symmetric except for the fact that $\mu_A$ and $\mu_B$
are different,
which constitutes a soft CP breaking.
Soft CP breaking causes spontaneous CP breaking,
embodied in the fact that $v_A \neq v_B$.\footnote{If we want
to avoid soft CP breaking,
we may alternatively add to the theory a real CP-odd scalar singlet.
Its vacuum expectation value breaks CP and induces $v_A \neq v_B$.}

There are four charged gauge bosons,
$W_A^\pm$,
with mass $(g v_A) / \sqrt{2}$,
and $W_B^\pm$,
with mass $(g v_B) / \sqrt{2}$.
I shall assume that $v_B \gg v_A$.
Then,
the $W^\pm_A$ are identified with the known $W^\pm$ gauge bosons,
while the $W^\pm_B$ mediate extra charged weak interactions.

Instead of $g$ and $g^\prime$ we may use the angle $\theta_w$,
defined by
\ba
s_w = \sin \theta_w &=& - \frac{g^\prime}{\sqrt{g^2 + 2 g^{\prime 2}}},
\nonumber\\
c_w = \cos \theta_w &=& \sqrt{\frac{g^2 + g^{\prime 2}}{g^2
+ 2 g^{\prime 2}}},
\nonumber\\
\sqrt{c_w^2 - s_w^2} &=& \frac{g}{\sqrt{g^2 + 2 g^{\prime 2}}}, 
\ea
together with the electric-charge unit $e = g s_w$.
The neutral gauge bosons of SU(2)$_A$,
SU(2)$_B$ and U(1),
named $W_{3A}$,
$W_{3B}$ and $B$,
respectively,
may be rotated by means of an orthogonal matrix,
\be
\left( \begin{array}{c} W_{3A} \\ W_{3B} \\ B \end{array} \right)
=
\left( \begin{array}{ccc}
- s_w & c_w & 0 \\*[1mm]
- s_w & - \frac{s_w^2}{c_w} & \frac{1}{c_w} \sqrt{c_w^2 - s_w^2}
\\*[1mm]
\sqrt{c_w^2 - s_w^2} & \frac{s_w}{c_w} \sqrt{c_w^2 - s_w^2}  &
\frac{s_w}{c_w} 
\end{array} \right)
\left( \begin{array}{c} A \\ Z \\ X \end{array} \right),
\ee
to obtain the photon $A$ and two other gauge bosons $Z$ and $X$,
which however are not eigenstates of mass.
The neutral gauge couplings are
\ba
 &   & g ( W_{3A} T_{3A} + W_{3B} T_{3B} ) + g^\prime B Y
\nonumber \\*[1mm]
 & = & - e A Q
+ \frac{g}{c_w} Z ( T_{3A} - Q s_w^2 )
+ \frac{g \sqrt{c_w^2 - s_w^2}}{c_w} X
\left( T_{3B} - Y \frac{s_w^2}{c_w^2 - s_w^2} \right).
\ea
With the above definitions,
$Z$ has interactions
similar to the ones of the known $Z^0$ of the standard model.
However,
$Z$ mixes with $X$.
The mass terms of $Z$ and $X$ are
\be
\frac{g^2}{4 c_w^2}
\left( \begin{array}{cc} Z & X \end{array} \right)
\left( \begin{array}{cc}
v_A^2 & \frac{s_w^2}{\sqrt{c_w^2 - s_w^2}} v_A^2 \\*[3mm]
\frac{s_w^2}{\sqrt{c_w^2 - s_w^2}} v_A^2 &
\frac{c_w^4 v_B^2 + s_w^4 v_A^2}{c_w^2 - s_w^2}
\end{array} \right)
\left( \begin{array}{c} Z \\ X \end{array} \right).
\ee
As a consequence,
in the limit $v_B \gg v_A$ $Z$ becomes an eigenstate of mass
with mass equal to $ (g v_A) / (\sqrt{2} c_w) $,
bearing the same relationship to the mass of $W_A^\pm$
as in the standard model,
while $X$ has a much larger mass $ ( g v_B c_w) /
\sqrt{2 (c_w^2 - s_w^2)}$.
For finite $v_B$,
$Z$ mixes with $X$,
the mixing angle being of order $v_A^2 / v_B^2$.

The fermions of the SM are duplicated
with new fermions in doublets or singlets of the SU(2)$_B$ gauge group.
The fermion spectrum is
\ba
q_L: (\two, \one)_{1/6}, \
p_R: (\one, \one)_{2/3}, \
n_R: (\one, \one)_{-1/3}, & &
\nonumber\\
l_L: (\two, \one)_{-1/2}, \
e_R: (\one, \one)_{-1}, & &
\nonumber\\
Q_L: (\one, \two)_{1/6}, \
P_R: (\one, \one)_{2/3}, \
N_R: (\one, \one)_{-1/3}, & &
\nonumber\\
L_L: (\one, \two)_{-1/2}, \
E_R: (\one, \one)_{-1}. & &
\ea
Here,
\be
q_L = \left( \begin{array}{c} p_L \\ n_L \end{array} \right), \
l_L = \left( \begin{array}{c} \nu_L \\ e_L \end{array} \right), \
Q_L = \left( \begin{array}{c} P_L \\ N_L \end{array} \right), \
L_L = \left( \begin{array}{c} \eta_L \\ E_L \end{array} \right).
\ee
Notice that the hypercharge assignments are the same as
in the standard model.
Also notice that the extra right-handed fermions
have the same quantum numbers as the usual right-handed fermions.
The CP transformation interchanges upper-case with lower-case fields;
thus,
\be
p_L \stackrel{\rm CP}{\rightarrow} \gamma^0 C \overline{P_L}^T, \
n_L \stackrel{\rm CP}{\rightarrow} \gamma^0 C \overline{N_L}^T, \
p_R \stackrel{\rm CP}{\rightarrow} \gamma^0 C \overline{P_R}^T, \
n_R \stackrel{\rm CP}{\rightarrow} \gamma^0 C \overline{N_R}^T,
\ee
and so on.
The Yukawa interactions are
\ba
{\cal L}_Y
& = &
- \overline{q_L} \phi_A (\Gamma n_R + \Gamma^\prime N_R)
- \overline{q_L} \tilde{\phi}_A (\Delta p_R + \Delta^\prime P_R)
\no
&   &
- \overline{l_L} \phi_A (\Lambda e_R + \Lambda^\prime E_R)
\no
&   &
- \overline{Q_L} \phi_B (\Gamma^\ast N_R + \Gamma^{\prime \ast} n_R)
- \overline{Q_L} \tilde{\phi}_B (\Delta^\ast P_R
+ \Delta^{\prime \ast} p_R)
\no
&   &
- \overline{L_L} \phi_B (\Lambda^\ast E_R + \Lambda^{\prime \ast} e_R)
+ \mbox{\rm H.c.},
\ea
where $\Gamma$,
$\Gamma^\prime$,
$\Delta$,
$\Delta^\prime$,
$\Lambda$ and $\Lambda^\prime$
are arbitrary complex $n_g \times n_g$ matrices in generation space
($n_g = 3$ is the number of generations).
Therefore,
the quark mass terms are
\ba
& &
- \left( \begin{array}{cc}
\overline{n_L} & \overline{N_L}
\end{array} \right)
\left( \begin{array}{cc}
v_A \Gamma & v_A \Gamma^\prime \\ 
v_B \Gamma^{\prime \ast} & v_B \Gamma^\ast \\ 
\end{array} \right)
\left( \begin{array}{c}
n_R \\ N_R
\end{array} \right)
\no
& &
- \left( \begin{array}{cc}
\overline{p_L} & \overline{P_L}
\end{array} \right)
\left( \begin{array}{cc}
v_A \Delta & v_A \Delta^\prime \\
v_B \Delta^{\prime \ast} & v_B \Delta^\ast \\
\end{array} \right)
\left( \begin{array}{c}
p_R \\ P_R
\end{array} \right)
+ \mbox{\rm H.c.}.
\ea
The determinants of the quark mass matrices are real.
Together with the assumption of CP invariance of the Lagrangian,
this implies that strong CP violation is absent at tree level.

The bi-diagonalization of the mass matrices follows the usual lines.
Let
\be
\left( \begin{array}{c} p_L \\ P_L \end{array} \right)
=
\left( \begin{array}{c} X_{u} \\ Y_{u} \end{array} \right)
\left( u_L \right) \ \mbox{\rm and} \
\left( \begin{array}{c} n_L \\ N_L \end{array} \right)
=
\left( \begin{array}{c} X_{d} \\ Y_{d} \end{array} \right)
\left( d_L \right).
\ee
There are $(2 n_g)$ quarks $u_L$
and $(2 n_g)$ quarks $d_L$.
The bi-diagonalization matrices,
which have been written in a convenient block form,
are $(2 n_g) \times (2 n_g)$ unitary.
The matrices $X_u$,
$Y_u$,
$X_d$ and $Y_d$ and $n_g \times (2 n_g)$.
The charged weak interactions of the quarks are given by
\be
\frac{g}{\sqrt{2}} \overline{u_L} \gamma^\mu
\left( W^+_{A \mu} V_A + W^+_{B \mu} V_B \right) d_L
+ \mbox{\rm H.c.},
\ee
where $V_A = X_{u}^\dagger X_{d}$ and $V_B = Y_{u}^\dagger Y_{d}$
are the generalized Cabibbo--Kobayashi--Maskawa (CKM) matrices.
I define
the $(2 n_g) \times (2 n_g)$ hermitian idempotent matrices
\be
H_u = X_{u}^\dagger X_{u} = V_A V_A^\dagger \ \mbox{\rm and} \
H_d = X_{d}^\dagger X_{d} = V_A^\dagger V_A.
\ee
Then,
$V_B V_B^\dagger = Y_{u}^\dagger Y_{u} = 1 - H_u$
and $V_B^\dagger V_B = Y_{d}^\dagger Y_{d} = 1 - H_d$
are hermitian and idempotent too.
The neutral gauge interactions of the quarks are given by
\ba
 & &
\overline{u} \gamma_\mu
\left[
- \frac{2}{3} e A^\mu
+ \frac{g}{c_w} Z^\mu
\left( \frac{1}{2} H_u \gamma_L - \frac{2}{3} s_w^2 \right)
\right.
\no
 & &
\left.
+ \frac{g \sqrt{c_w^2 - s_w^2}}{c_w} X^\mu
\left( - \frac{2}{3} \frac{s_w^2}{c_w^2 - s_w^2}
+ \frac{1}{2} \frac{c_w^2}{c_w^2 - s_w^2} \gamma_L
- \frac{1}{2} H_u \gamma_L \right)
\right]
u
\no
 & &
+ \overline{d} \gamma_\mu
\left[
\frac{1}{3} e A^\mu
+ \frac{g}{c_w} Z^\mu
\left( \frac{1}{3} s_w^2 - \frac{1}{2} H_d \gamma_L \right)
\right.
\no
 & &
\left.
+ \frac{g \sqrt{c_w^2 - s_w^2}}{c_w} X^\mu
\left( \frac{1}{3} \frac{s_w^2}{c_w^2 - s_w^2}
- \frac{1}{2} \frac{c_w^2}{c_w^2 - s_w^2} \gamma_L
+ \frac{1}{2} H_d \gamma_L \right)
\right]
d,
\ea
where $\gamma_L = (1 - \gamma_5) / 2$.
Denoting $M_d = \mbox{\rm diag} (m_d, m_s, m_b, ...)$
the diagonal real matrix of the masses of the $(2 n_g)$
down-type quarks,
and similarly $M_u = \mbox{\rm diag} (m_u, m_c, m_t, ...)$
for the up-type quarks,
the Yukawa interactions of the quarks are given by
\ba
 & - & \frac{H_A}{\sqrt{2} v_A}
\left( \overline{d_L} H_d M_d d_R + \overline{u_L} H_u M_u u_R \right)
\no
 & - & \frac{H_B}{\sqrt{2} v_B}
\left[ \overline{d_L} (1 - H_d) M_d d_R
+ \overline{u_L} (1 - H_u) M_u u_R \right]
+ \mbox{\rm H.c.}.
\ea

We may now see that strong CP violation still vanishes
at one-loop level.
Indeed,
at one-loop level the only diagrams
which may generate complex mass terms for the quarks
have in the loop either one of the two physical scalar particles,
or one of the two neutral massive gauge bosons.
Those diagrams are real except for the presence
of the matrices $H_d$ and $H_u$ in the couplings.
Even though they generate complex off-diagonal mass terms,
the diagonal mass terms are still real,
because of the hermiticity of $H_d$ and $H_u$.
As a consequence,
the one-loop contribution to strong CP violation \cite{ellis},
which is $ \arg \mbox{\rm tr} (\Sigma M^{-1}) $
($\Sigma$ being the one-loop-level contribution to the quark mass matrix
and $M$ the diagonal quark mass matrix at tree level),
vanishes.

It is instructive to check that CP is indeed conserved when $v_B = v_A$.
In that case,
the quark mass matrices are of the form
(see eq.~14)
\be
\left( \begin{array}{cc} M & N \\ N^\ast & M^\ast \end{array} \right),
\ee
$M$ and $N$ being complex $n_g \times n_g$ matrices.
It is easy to demonstrate that a choice
of the fields eigenstates of mass,
including in particular a phase convention for those fields,
exists such that the unitary diagonalizing matrices
have $Y_d = X_d^\ast$ and $Y_u = X_u^\ast$.
For this choice of physical quark fields $V_B = V_A^\ast$.
On the other hand,
when $v_B = v_A$ the masses of the $W_B^\pm$ and of the $W_A^\pm$
are equal.\footnote{It must be stressed however that,
if both $M$ and $N$ are non-zero matrices,
the quarks will in principle all have different masses.
This fact does not affect the reasoning.}
It follows that,
then,
the two charged weak interactions have the same strength
but opposite CP-violating effects,
which implies that altogether CP is conserved in those interactions.

As for the neutral weak interactions,
when $v_B = v_A$
the mass matrix of $Z$ and $X$
(see eq.~9)
is such that one of its eigenstates couples to a real current
while the other eigenstate couples to a current of the form
\be
\overline{u_L} \gamma^\mu (2 H_u - 1) u_L -
\overline{d_L} \gamma^\mu (2 H_d - 1) d_L.
\ee
Now,
$ 2 H_u - 1 = X_u^\dagger X_u - Y_u^\dagger Y_u $
is an imaginary matrix because $Y_u = X_u^\ast$.
Similarly,
$2 H_d - 1$ is an imaginary matrix.
As a consequence,
this neutral current is purely imaginary,
which implies no CP violation.

Finally,
and in an analogous fashion,
it can be shown that when $v_B = v_A$
one of the physical scalars has real Yukawa couplings
while the other one has imaginary Yukawa couplings,
which means that there is no CP violation
in the Yukawa interactions either.

We conclude that CP is indeed conserved when $v_B = v_A$.

Let us consider how large $v_B$ should be.
All $2 n_g = 6$ neutrinos are massless in this model;
this is not a problem,
because the three $\eta_L$ neutrinos have $Q = T_{3A} = 0$,
and therefore they do not contribute to the width of the $Z^0$,
except for the fact that $Z^0$ contains a small admixture,
of order $v_A^2 / v_B^2$,
of $X$.
All other extra fermions contribute to the width of the $Z^0$
unless they are massive enough.
For $v_B \gg v_A$ we expect the masses of the new fermions
to be of order $v_B / v_A$ times the masses of the ordinary fermions.
This would lead us to guess that $v_B / v_A$
should be no smaller than $10^5$,
so that the heavy counterpart of the electron does not contribute
to the $Z^0$ width.
However,
the masses of the heavy fermions are not really $v_B / v_A$
times the masses of the ordinary fermions.
$v_B / v_A$ may easily be as low as $10^2$ or so.
Even for such a low $v_B / v_A$,
the $W_B^\pm$ would have a very high mass $\sim 8$ TeV,
and the admixture of $X$ in the $Z^0$ would be $\sim 10^{-4}$,
and thus unobservably small.

The main consequence of a finite $v_B / v_A$ is fermion mixing.
Indeed,
for $v_B \gg v_A$
the bi-diagonalization matrices acquire an approximate block form;
the $n_g \times n_g$ left blocks of $X_d$ and $X_u$,
and right blocks of $Y_d$ and $Y_u$,
become approximately unitary,
while the $n_g \times n_g$ right blocks of $X_d$ and $X_u$,
and left blocks of $Y_d$ and $Y_u$,
get suppressed by $v_A / v_B$.
Thus,
the CKM matrix $V_A$
has its $3 \times 3$ upper left-hand block approximately unitary,
with its other $3 \times 3$ blocks
suppressed by one or two powers of $v_A / v_B$.
Similarly,
the $6 \times 6$ hermitian matrices $H_u$ and $H_d$
have their upper left-hand $3 \times 3$ blocks
approximately equal to the unit matrix,
and all other $3 \times 3$ blocks suppressed.
This ensures that,
for large $v_B / v_A$,
approximate unitarity of the CKM matrix is recovered
and the $Z^0$ interactions with the fermions
are approximately diagonal
as they should.
For practical purposes,
this model is analogous to a model
which simultaneously has $n_g$ vectorlike isossinglet down-type quarks
and $n_g$ vectorlike isossinglet up-type quarks,
with the mixing between the normal quarks and the isossinglet quarks
suppressed by $v_A / v_B$.

It must however be stressed that,
if we want to totally evade the constraints on fermion mixing,
we may simply provide the model with an extra $Z_2$ symmetry
under which all new fermion fields change sign
and all other fields remain invariant.
This symmetry eliminates the Yukawa-coupling matrices $\Gamma^\prime$,
$\Delta^\prime$ and $\Lambda^\prime$,
thus ensuring the absence of fermion mixing.
A consequence of such a symmetry is that the masses of the new fermions
are then exactly proportional to the masses of the ordinary fermions,
the proportionality factor being $v_B / v_A$,
which would then certainly have to be larger that $10^5$.
My model would then be rather similar to the ones in ref.~\cite{prl},
except for the fact that CP symmetry is used here instead of P symmetry.

With that extra $Z_2$ symmetry strong CP violation
arises only at three-loop level \cite{ellis}
and is irrelevantly small.
On the other hand,
without the $Z_2$ symmetry the matrices $H_u$ and $H_d$
have non-vanishing off-diagonal matrix elements,
which allows for the generation
of strong CP violation already at two-loop level.
However,
that two-loop-level strong CP violation will have a suppression
$(v_A / v_B)^2$ \cite{BM} and,
if we reasonably assume that $v_B / v_A$ should at least be
$\sim 10^2$,
we conclude that
this two-loop contribution to strong CP violation is inoffensive too.
We may thus say that this model explains the smallness
of strong CP violation
by relating it to the observed absence
of flavour-changing neutral currents
and to the identity between the observed $Z^0$ width
and its standard-model prediction,
which both point out to a large value of $v_B / v_A$.

In conclusion,
I have presented a model in which CP is violated
as the result of a spontaneous breaking of CP of a new type.
My model shares with the left-right-symmetric model \cite{BM}
the feature that,
if the vacuum expectation value $v_B$ goes to infinity,
the theory becomes identical to the standard model
and no trace of the spontaneous nature of CP breaking
(in the left-right--symmetric model,
of P breaking)
remains.
My model solves the strong-CP problem by incorporating the idea
\cite{ellis}
that there must be some physics at higher energies
which cancels the tree-level contribution to the $\theta$ parameter.
My model avoids the much expanded scalar sector,
and in particular the uncomfortable phenomenon
of scalar-pseudoscalar mixing,
which are typical of the usual models of spontaneous CP breaking.

\vspace{5mm}

I thank Walter Grimus for reading the manuscript
and making useful criticisms.

\end{document}